\documentclass[aps,prd,floatfix,10pt,nofootinbib,showpacs,notitlepage,superscriptaddress]{revtex4-1} 
\usepackage{amsmath}
\usepackage{graphicx}  % needed for figures
\usepackage{dcolumn}   % needed for some tables
\usepackage{bm}        % for math
\usepackage{amssymb}   % for math

\usepackage{dsfont}
\usepackage{mathrsfs}
\usepackage{times}
\usepackage[colorlinks=true,linkcolor=blue,citecolor=blue,urlcolor=blue]{hyperref} 
\usepackage{gensymb} 

\usepackage{color}

\begin{document}
\title{Analogue Wormholes and Black Hole LASER Effect in Hydrodynamics}
\author{C\'edric Peloquin}
\affiliation{Universit\'e Fran\c{c}ois Rabelais de Tours, 60 Rue du Plat d'Etain, 37000 Tours, France}
\author{L\'eo-Paul Euv\'e}
\affiliation{Pprime Institute, UPR 3346, CNRS - Universit\'e de Poitiers - ISAE ENSMA,
11 Boulevard Marie et Pierre Curie, T\'el\'eport 2, BP 30179, 86962 Futuroscope Cedex, France}
\author{Thomas Philbin}
\affiliation{Physics and Astronomy Department, University of Exeter,
Stocker Road, Exeter EX4 4QL, United Kingdom}
\author{Germain Rousseaux}
\affiliation{Pprime Institute, UPR 3346, CNRS - Universit\'e de Poitiers - ISAE ENSMA,
11 Boulevard Marie et Pierre Curie, T\'el\'eport 2, BP 30179, 86962 Futuroscope Cedex, France}

%\date{\today}

\begin{abstract}
We numerically study water wave packets on a spatially varying counter-current in the presence of surface tension. Depending on the details of the velocity profile, we show that traversable and bi-directional analogue wormholes exist in fluid mechanics. The limitations on traversability of wormholes in general relativity are absent here because of the dispersion of water waves and the ability to form flow profiles that are not solutions of Einstein's equations. We observe that negative energy can be trapped between analogue horizons forming a LASER-like cavity. Six horizons are involved in the trapping cavity because of the existence of two dispersive scales, in contrast to previous treatments which considered two horizons and one dispersive scale. 
\end{abstract}

\maketitle

\section{Introduction}

A wormhole is a connexion between separate parts of the universe featuring a black hole (or gravitational drain) and its time-reversed partner, a white hole (or gravitational fountain). 
The Einstein-Rosen bridge of General Relativity is a wormhole with one black and one white hole horizon but it is not traversable (even in one direction) because it closes up too quickly \cite{ER, FW}. Thorne et al. hypothesized the existence of exotic matter possessing negative energy in order to stabilized wormholes that can become two-way tunnels through space-time \cite{MTY}.  Following these authors, one may hypothesize that an "exotic" wormhole may be traversable and bi-directional. Strictly speaking, an "exotic" wormhole would no longer feature horizons in the sense of classical General Relativity. 

Analogue Gravity studies the similarities at the level of kinematics between the propagation of light in curved space-times and the propagation of waves in moving condensed matter media \cite{BLV, Como}: the Hawking radiation of a horizon is currently the archetypal example of astrophysical phenomenon that is studied by condensed matter groups in the laboratory. It is well known that the analogy breaks down at the level of dynamics since the dynamical equations are different : Einstein's equations for space-time and say, Navier-Stokes equations for classical fluid analogues. Nevertheless, Sch\"utzhold and Unruh demonstrated that water waves propagating on a flow current can be seen as a gravitational analogue for the propagation of light in a curved space-time in the long wavelength regime corresponding to shallow gravity waves \cite{SU, Como}. This analogue system is currently under experimental scrutiny \cite{NJP08, NJP10, Silke, Leo}. In this condensed matter system, the velocity of light is mimicked by the velocity of non-dispersive long gravity waves in water with speed $|c|=\sqrt{gh}$ where $h$ is the water depth \cite{SU, Como}. The horizon is the place where the velocity of the flow $U(x)>0$ (the flow is positive when going to the right) is equal in modulus to the velocity of the wave $c(x)<0$ known as the critical condition in fluid mechanics and where $x$ is the position along the water channel. The flow current when varying in space induces an effective curved space-time for the hydrodynamic waves propagating on top of it. For example, the Schwarzschild spherical geometry is analogous to a flow with a current given by $U(r)=|c|\sqrt{r_s/r}$ where $r_S$ is the Schwarzschild radius namely a constant which depends only on the mass of the black hole and the gravitational constant. Obviously, the "flow current" for the Schwarzschild geometry goes to zero at infinity. Analogue Gravity is, in a sense, more general that General Relativity since it encompasses as many velocity profiles as experimentalists in condensed matter can design. From the point of view of a relativist, the analogy is not exact even if the flow goes to zero because $U(x)$ does not reproduce exactly the flow profile of the Schwarzschild space-time even in that case. From the point of view of Analogue Gravity, one can generate any flow profile and look for the common features between the experiments in condensed matter Physics and the relativistic predictions which are narrower since they are derived in the particular velocity profile corresponding to the Schwarzschild space-time. In water wave Physics, the flow is generated with a pump in a water channel and the variations are induced either by a changing depth or width. Hence, using a bump one can generate easily sub-critical ($U<|c|$) and super-critical ($U>|c|$) regions. In the experiments performed so far, either the flow vanishes far from the bump \cite{NJP08, NJP10} or it becomes uniform \cite{Silke, Leo}. Strictly speaking, the flat space-time would correspond either to a vanishing current velocity region or a constant flow region whereas a curved space-time corresponds to a region with an inhomogeneous current in space.

We can possibly design "exotic wormholes" by looking at "analogue wormholes" namely pairs of horizons which are stable in Analogue Gravity contrary to General Relativity since the dynamical equations are not the same. We will look in this work at the fate of a wave-packet propagating into an analogue wormhole in Hydrodynamics. By looking first at the non-dispersive case, we will recover similar behavior to that in General Relativity. Indeed, a major interest of the Analogue Gravity program is the inclusion of dispersion which removes the infinite blue-shifting of waves close to the horizon of a black/white hole, avoiding the so-called trans-Planckian problem \cite{Unruh, Jacobson}. The Hawking radiation of horizons has been shown to be robust in certain regimes even in the presence of dispersion \cite{BLV, Como,Scott}. Then, the existence of bi-directional and stable "analogue" wormholes in Hydrodynamics will be demonstrated. Their stability is due to dynamical effects (the Navier-Stokes equation allows such stationary flows to exist), while their bi-directionality is due to dispersive effects.

A straightforward consequence will be the appearance of an analogue Black Hole LASER effect between pairs of horizons playing the role of a LASER cavity in a super-critical region. The latter effect will be implemented in water-wave Physics with an anomalous dispersion relation following Sch\"utzhold and Unruh \cite{SU} who described a version with normal dispersion. The black-hole LASER effect was theoretically predicted for the first time by Corley and Jacobson \cite{CJ} and is further discussed in \cite{Philbin}. We note also its numerical demonstration in optics \cite{Faccio} and its recent experimental observation in a Bose-Einstein Condensate by Steinhauer \cite{Steinhauer}. Because the dispersion is anomalous with Bogoliubov phonons, self-amplification of the Hawking radiation will lead to a lasing effect.

\section{Non-Dispersive Wormholes in Hydrodynamics}

\begin{figure}[!htbp]
\includegraphics[scale=0.6]{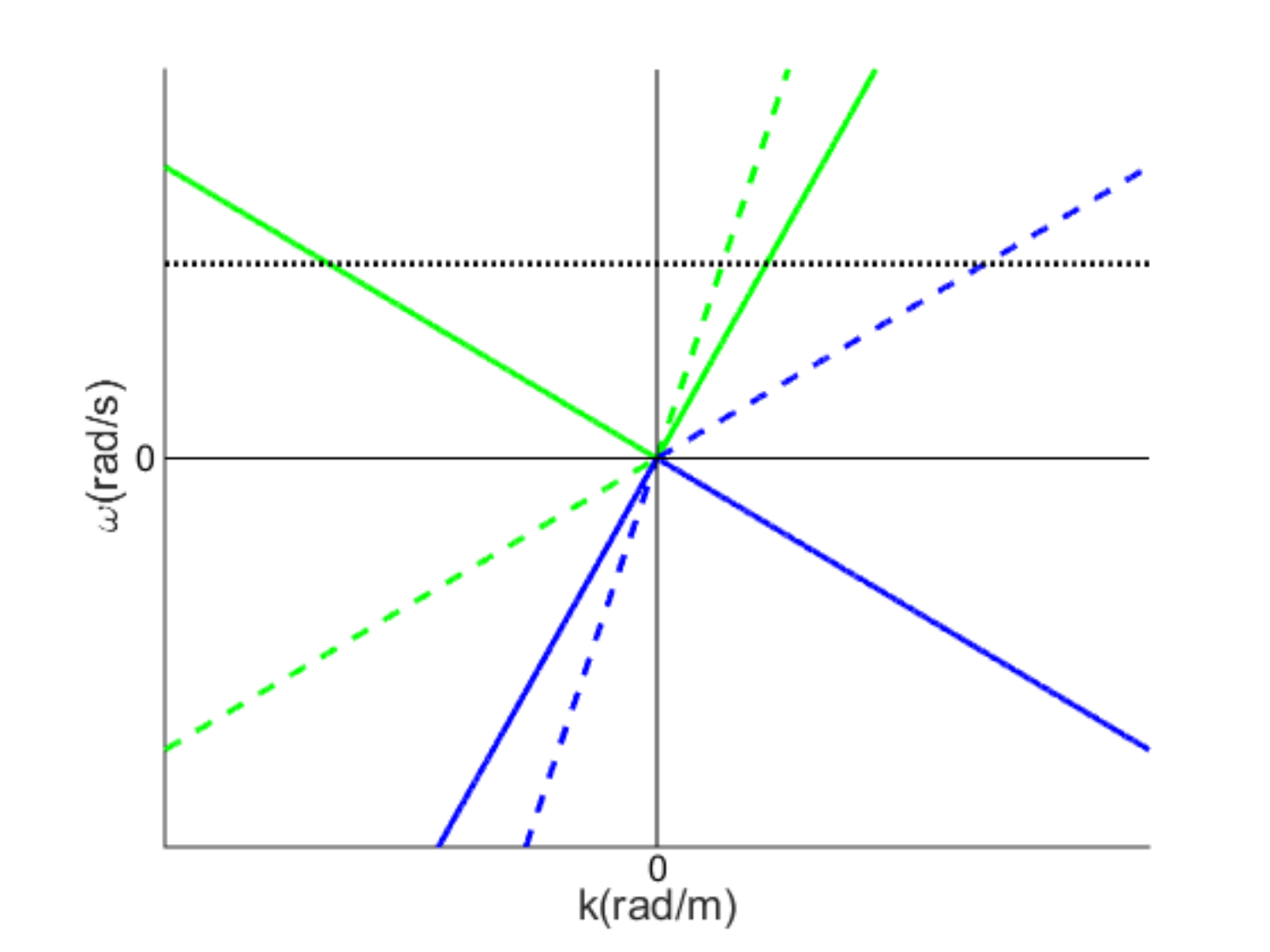}
\caption{Plots of the dispersion relation in the non-dispersive limit for the black hole case ($\omega=(U\pm c)k$, $U>0$) for a subcritical region where $U=0.5\cdot |c|$ (continuous straight lines) and for a supercritical region where $U=1.5\cdot |c|$ (dashed lines). The positive relative frequency $\omega'=\omega -Uk=\pm ck'=\pm ck$ corresponds to the green color (the ' corresponds to the flow current frame of reference); the negative relative frequency is in blue. The conserved frequency of an incident wave as generated by a wave-maker is the horizontal line in dotted black.}
\label{relativist_rd}
\end{figure}

The dispersion relation in the dispersion-less regime of Analogue Gravity is : $\omega=(U\pm c)k$ where $\omega$ is the angular frequency and $k$ is the longitudinal wavenumber (see the Figure \ref{relativist_rd}). In water waves Physics, $U$ is the flow current whereas $c$ is the water wave velocity. The reader is referred to the review chapter in \cite{Como} on the dispersive effects met in water waves Physics. In the super-critical region (for example in the interior of a black hole in General Relativity), negative energy waves with negative relative frequency ($\omega -Uk<0$) are present and they are the analogue of anti-particles in Quantum Field Theory whereas the positive relative frequency solutions ($\omega -Uk>0$) of the dispersion relation are the analogue of particles and they only live in the sub-critical region (namely outside of the black/white hole) if the regime is dispersion-less. The domain of existence of analogue particles/antiparticles will change due to the presence of dispersion and will no longer be separated necessarily by the horizon as in the dispersion-less regime.

\begin{figure}[!htbp]
\includegraphics[scale=0.8]{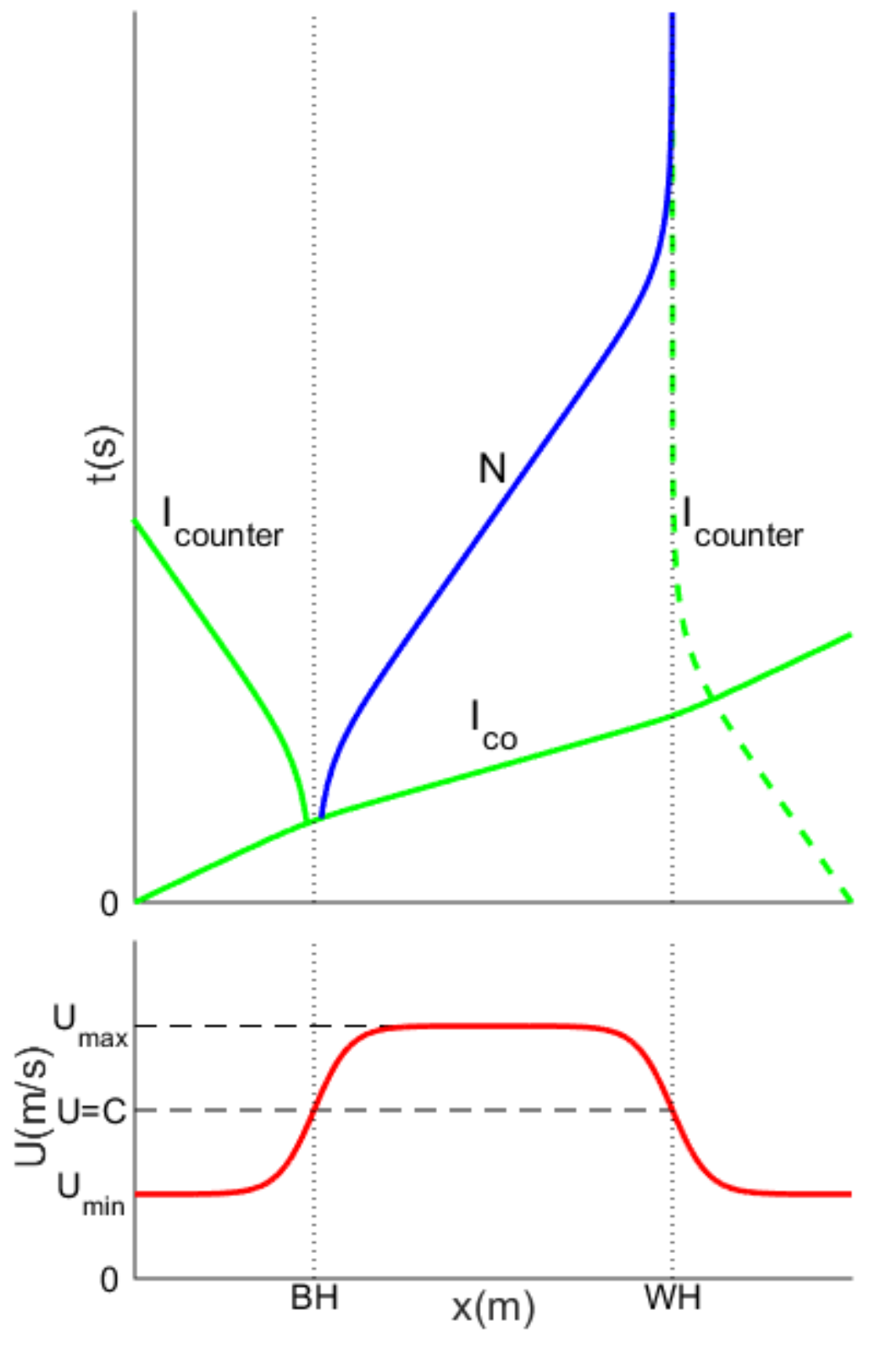}
\includegraphics[scale=0.8]{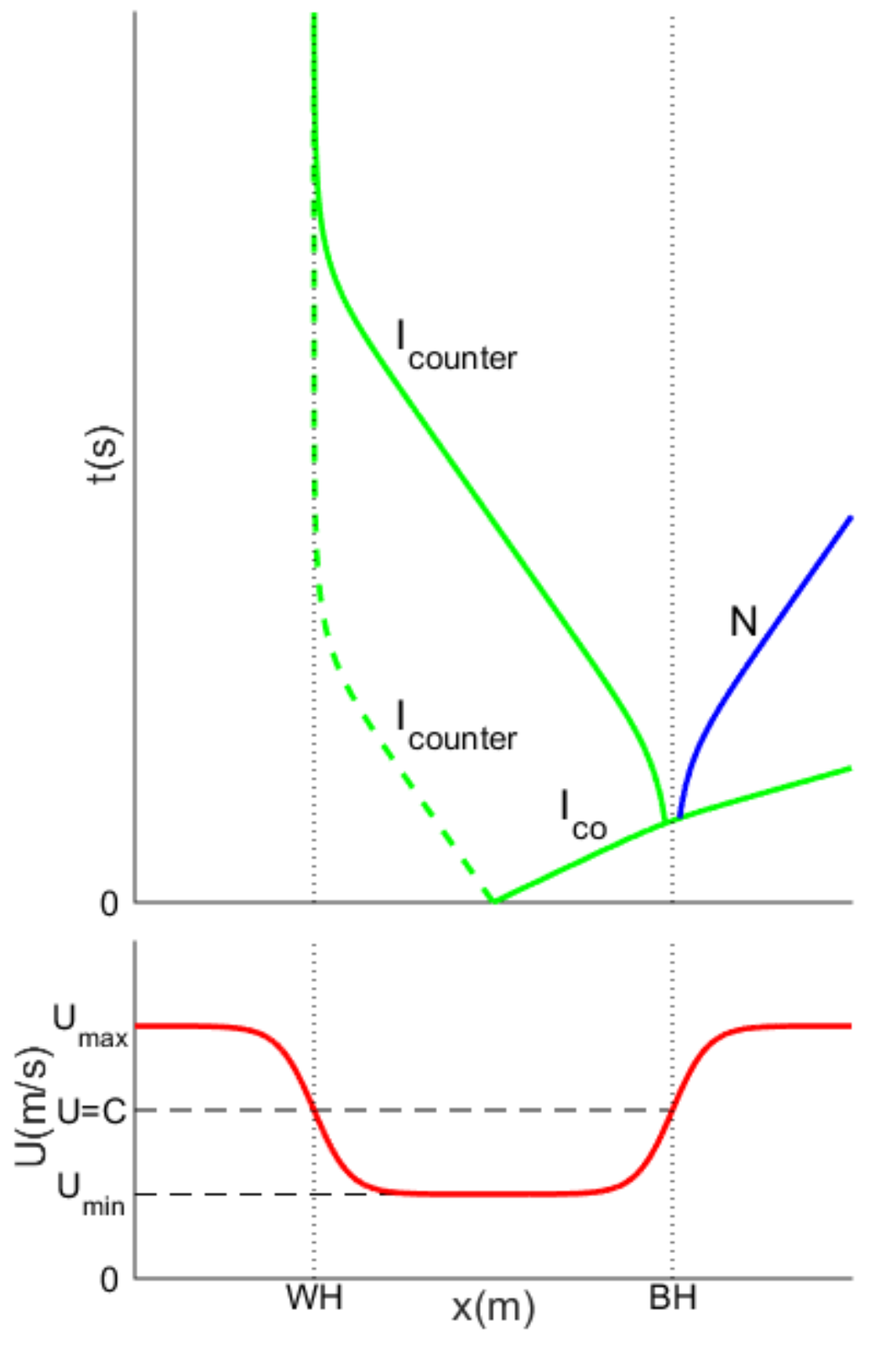}
\caption{Left bottom: velocity field for a bump geometry. Right bottom: velocity field for a trough geometry. Left and right top: scheme of the associated space-time diagrams. For both cases, we set $U_{min}=0.5\cdot c$ and $U_{max}=1.5\cdot c$.}
\label{relativist}
\end{figure}

We can create a pair of horizons in a water channel either by changing twice the depth or the width of its cross section. For example, if the current flows over a bump, the current can reach first a black hole horizon in the accelerated region on the ascending slope ($dU/dx>0$ and $U=|c|$) and then a white hole horizon in the decelerated region on the descending slope ($dU/dx<0$ and $U=|c|$). Hence, in the inter-horizons region,  the flow is supercritical in the long wavelength approximation whereas the flow is subcritical on both sides of the bump (see the Figure \ref{relativist} left). Inversely, if the current flows over a trough, it will generate first a white horizon and then a black horizon. In this configuration the flow is subcritical between both horizons whereas it is supercritical outside (see the Figure \ref{relativist} right). For both the bump and the trough geometries and depending on the shape of the bump (for example, by including a flat plate on the top of it), we can reach a constant flow region of a finite extent where the flow is supercritical and where dispersion-less waves can only propagate in a co-current direction so are compelled to be expelled from this supercritical black/white hole region.

Contrary to General Relativity, a wormhole in Hydrodynamics would be stable and stationary provided the flow current is constant in time. As in General Relativity, an analogue wormhole in Hydrodynamics would be unidirectional if the medium is dispersion-less since an incoming wave-packet willing to enter into the analogue wormhole and generated in the sub-critical region outside the black hole (namely in the flat space-time region) can only propagate in the co-current direction by entering first into the black hole through the black horizon, then it would proceed in the supercritical region following the current, before escaping the white hole through the white horizon where he would reach a new flat space-time region ($I_{co}$ in the Figure \ref{relativist} left top in continuous line). This case will be denoted as - Black$\rightarrow$White - in the rest of the paper where continuous lines refer to wave packets and their mode-converted partners generated in co-current whereas dashed lines refer to the ones generated in counter-current. In the dispersion-less regime, the only direction of propagation which is permitted is - Black$\rightarrow$White - (co-current) : the incident wave entering into the analogue black hole is just refracted at the black horizon and at the white one by the velocity gradient. A larger velocity gradient ($\frac{\partial u}{\partial x}$) would be able to convert a part of the incident wave into a negative mode in the supercritical region and a positive counter current mode in the sub-critical region near to the black hole. But this negative mode would suffer an infinite blue-shifting when arriving on the white horizon (see the continuous lines in the Figure \ref{relativist} left top). Hence, due to the flow gradient (analogue to the tidal forces at the horizon), mode conversion happens (in continuous lines) : $I_{co}$ is converted partly into positive energy waves $I_{counter}$ and negative energy waves $N$ (see the Figure \ref{relativist} left top in continuous line). Conversely, the reverse travel of a wave-packet (starting in the subcritical region on the right of the bump in the Figure \ref{relativist} left that is outside the white hole region) through the very same wormhole in the direction - White$\rightarrow$Black - namely against the counter-current, would be impossible due to the infinite blue-shifting at the white hole horizon ($I_{counter}$ in the Figure \ref{relativist} left top in dashed line). This behavior is similar to the trans-Planckian problem in General Relativity ($I_{counter}$ in the Figure \ref{relativist} left top in dashed line) since the wavelength goes to zero at the horizon because $k=\omega/(U+c)$.

A similar phenomenon appears in the case of a wave-packet sent in the inter-horizon region in a trough velocity field (see the Figure \ref{relativist} right bottom). The incident wave-packet $I_{counter}$ going against the current is infinitely blue-shifted at the white horizon (see the Figure \ref{relativist} right top in dashed line). The incident mode $I_{co}$ in co-current goes through the black horizon, converts to a negative mode $N$ in the super-critical region and a positive counter-current mode in the sub-critical region $I_{counter}$ which is also characterized by a continuous line because it comes from the conversion of the wave-packet $I_{co}$. To resume, we separate the two cases (continuous and dashed) by the direction in which the wave packet is sent initially. In all the cases and without dispersion, the pathology of the infinite blue-shifting is present at the horizons as is well-known in Analogue Gravity \cite{BLV, Scott}.

\section{The Effect of Dispersion in Hydrodynamics}

A cascade of dispersive scales can be encountered in Analogue Gravity to deal with the trans-Planckian problem (see the chapter by Chaline et al. in \cite{Como}). In water wave Physics, the first obvious scale is the water depth $h$ \cite{Como}. When the wavelength gets shorter, the effect of surface tension becomes relevant \cite{Como}. Rousseaux et al. studied the effect of the capillary length on the blocking of water waves and they showed how it can play the role of a cut-off on analogue Hawking radiation \cite{NJP10, Como}. Following the experimental observation of Badulin et al. \cite{Badulin} and the calculations of Trulsen and Mei \cite{TM}, they provide a parameter space diagram where the blocking velocity of waves at horizons is a function of their incoming period for a white hole configuration (an incoming wave starting outside of the white hole horizon in the sub-critical region propagates in the direction opposite to the counter-current and it may be blocked and mode-converted). Rousseaux et al. characterized the different mode conversions featuring six wave-numbers and three horizons (white, blue and negative) described by the dispersion relation of water waves \cite{Dingemans, Como} ($\rho=1000kg.m^{-3}$ and $\gamma=0.074N.m^{-1}$ for water):
\begin{equation}
\omega=Uk\pm\sqrt{\left(gk+\dfrac{\gamma}{\rho}k^{3} \right)\tanh(kh)}.
\label{disp_G_flow}
\end{equation}
The latter may be derived from the generalized Unruh equation including both the dispersive effects of the water depth and surface tension \cite{NJP10, Como}:
\begin{equation}
(\partial_{t}+\partial_{x}U)(\partial_{t}+U\partial_{x})\phi = i \left(g\partial_{x} - \dfrac{\gamma}{\rho}\partial^{3}_{x} \right) \tanh(-ih\partial_{x})\phi.
\label{Unruh equation}
\end{equation}
Its long wavelength limit describes the propagation of light in curved space-time provided the velocity of light corresponds to the velocity of long gravity waves $c=\sqrt{gh}$ \cite{SU, Como}.

If we introduce the two dispersive lengths into the dispersion relation (the water depth $h$ and the capillary length $l_c=\sqrt{\frac{\gamma}{\rho g}}=2.7mm$ in water), we get:
\begin{equation}
(\omega-Uk)^2=\left(gk+\dfrac{\gamma}{\rho}k^{3} \right)\tanh(kh)=\left(gk+gl_c^{2}k^{3} \right)\tanh(kh)
\label{disp_G_flow2}
\end{equation}
If we approximate the dispersion relation by a Taylor series up to the third order and assuming $kh<<1$, we obtain:
\begin{equation}
(\omega-Uk)^2 \simeq ghk^2+(gl_c^2h-\frac{gh^3}{3})k^4.
\label{disp_G_flow2_limited_dvp}
\end{equation}
Hence, the water depth induces a dispersive correction in $-k^4$ whereas the capillary length induces a correction in $+k^4$ both for shallow depths \cite{SU, Como}. The latter positive correction is reminiscent of the Bose-Einstein Condensate scenario with Bogolyubov phonons \cite{Steinhauer} that can be tested with a circular jump experiment in Hydrodynamics \cite{Gil}. We will now study the fate of a wave-packet that enters into a white hole due to the two dispersive lengths and a double-bouncing scenario. To demonstrate the bi-directional nature of a  wormhole in Hydrodynamics, we will also study the entrance into a black hole which is the case usually discussed in General Relativity.

\section{General settings}

\subsection{Wavepacket}
\begin{figure}[!htbp]
\includegraphics[scale=0.65]{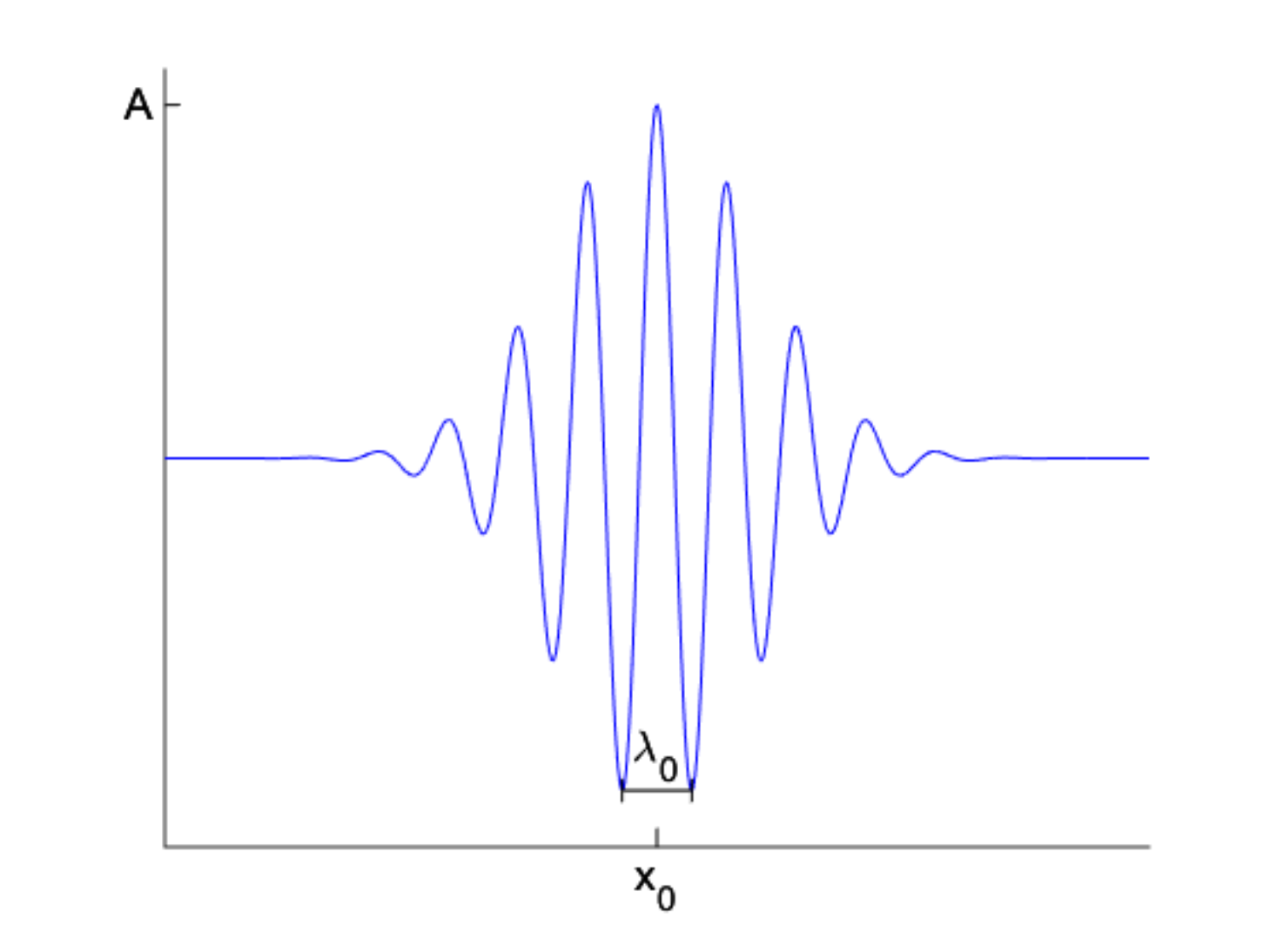}
\caption{The wave-packet as described by the equation (\ref{Gaussian_Wave_Packet}) used in all simulations with $\sigma=\lambda_0$.}
\label{wavepacket}
\end{figure}

We send on a flow a Gaussian wave-packet (see the Figure \ref{wavepacket}) whose mathematical expression is:
\begin{equation}
a(x)=A~e^{i \cdot k_{0}(x+x_{0})}~e^{-\dfrac{(x-x_{0})^{2}}{4\sigma^2}}
\label{Gaussian_Wave_Packet}
\end{equation}
where $A$ is its amplitude fixed to 1 in arbitrary units since we are in the linear approximation corresponding to small perturbations of the free surface. $k_{0}$ corresponds to the central wavenumber in the spectral representation of the wave-packet. $\sigma$ is a parameter controlling the initial width of the wave-packet. $x_{0}$ is the initial position of the wave-packet center.

\subsection{Velocity fields}

\begin{figure}[!htbp]
\includegraphics[scale=0.6]{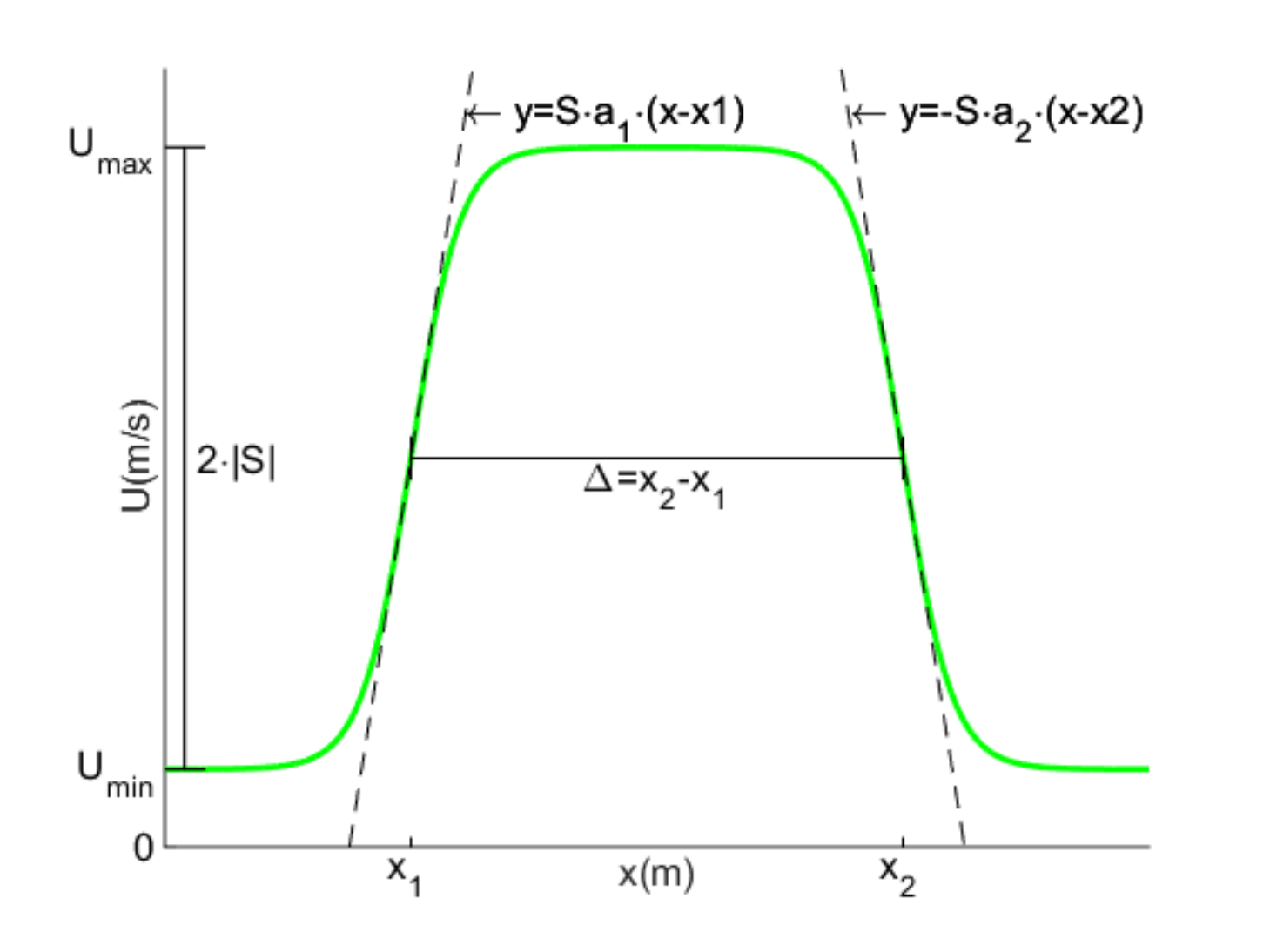}
\includegraphics[scale=0.6]{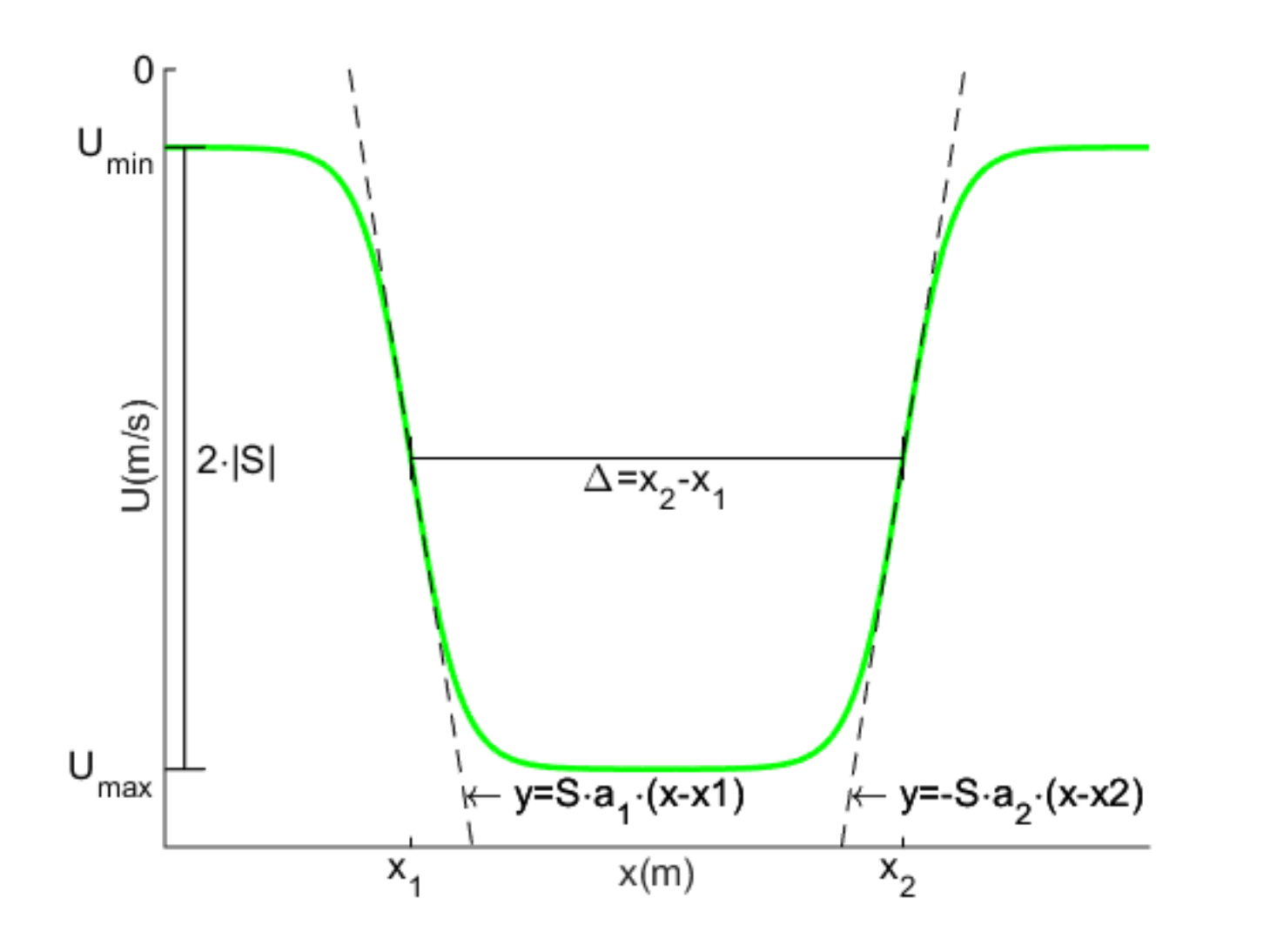}
\caption{Left: velocity field for the travel - Black$\rightarrow$White - (co-current $U>0$). Right: velocity field for the travel - White$\rightarrow$Black - (counter-current $U<0$).}
\label{velocity_field}
\end{figure}

As with previous simulations on the interaction of a current and a wave packet, we take as a model flow a hyperbolic tangent with two assumed non-dispersive horizons (white and black) separated by a given distance (see the Figure \ref{velocity_field}):
\begin{equation}
U(x)=U_{min}+S(\tanh(a_{1} \cdot (x-x_1)-\tanh(a_{2}(x-x_2)))
\label{Tanh}
\end{equation}
where $|U_{min}|$ is the minimal velocity far from the region where the current changes significantly, $a_{1}$ is the slope on the left side of the velocity profile and $a_{2}$ is its right counterpart. $\Delta=x_2-x_1$ is the distance between the two maxima of the velocity profile gradient. We recall that the gradient of the velocity profile taken at the non-dispersive horizon in Analogue Gravity is the analogue of the surface gravity in General Relativity and that it controls the mode conversion intensity \cite{BLV}. $S$ is the step velocity between the top and the bottom of the velocity profile: $U_{min}+2S=U_{max}$ is the maximum velocity reached by the fluid. The maximum gradients are, (for $i=1$ or $2$):
\begin{equation}
\left. \dfrac{\partial U}{\partial x} \right| _{x=x_i} = \pm a_{i} \cdot S
\end{equation}
The water depth is assumed constant in contrast to the velocity which changes with position, as if the width of the water channel were changing albeit with a fixed depth.

\subsection{Horizons and solutions}

\begin{figure}[!htbp]
\includegraphics[scale=0.6]{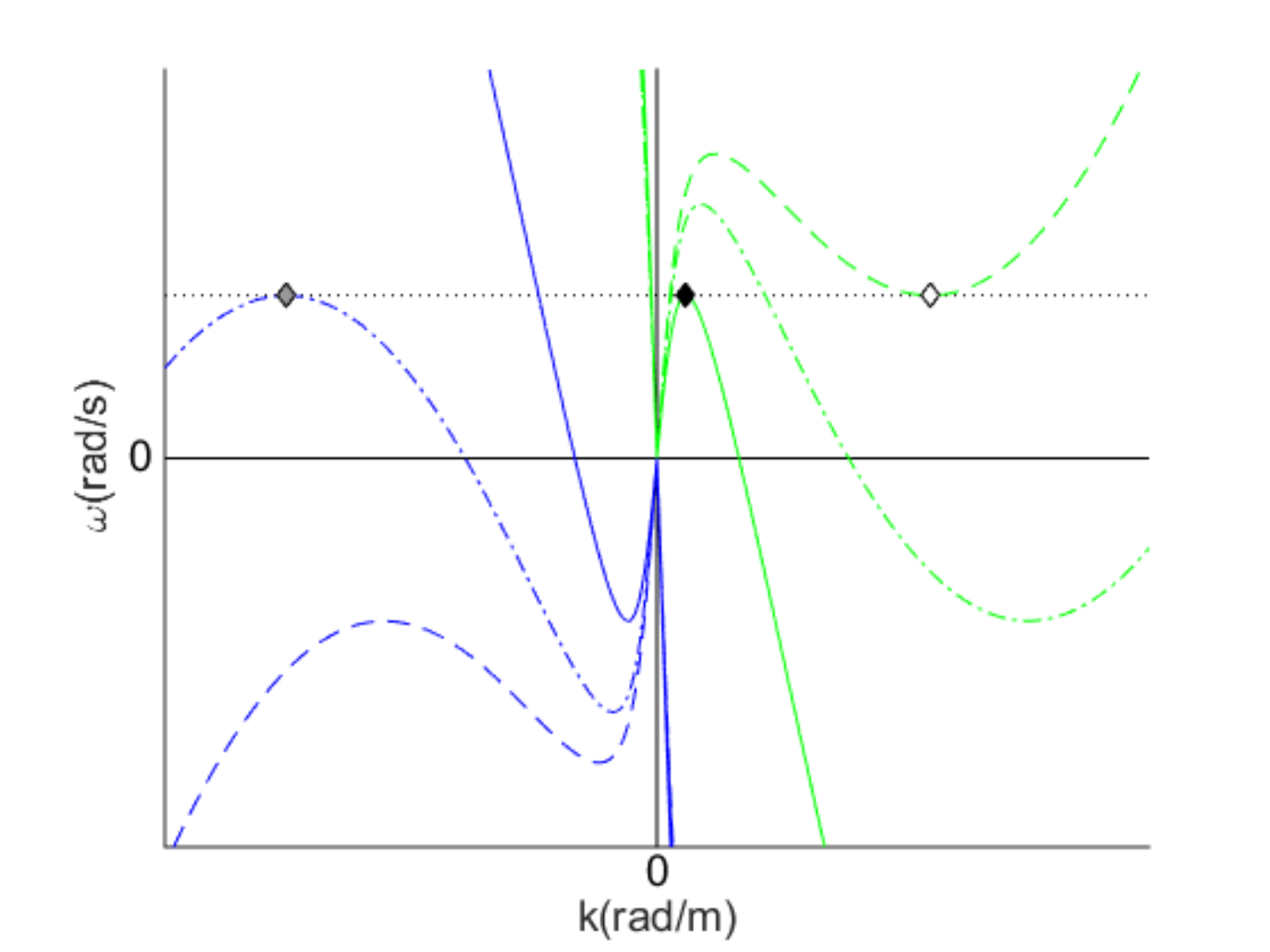}
\includegraphics[scale=0.6]{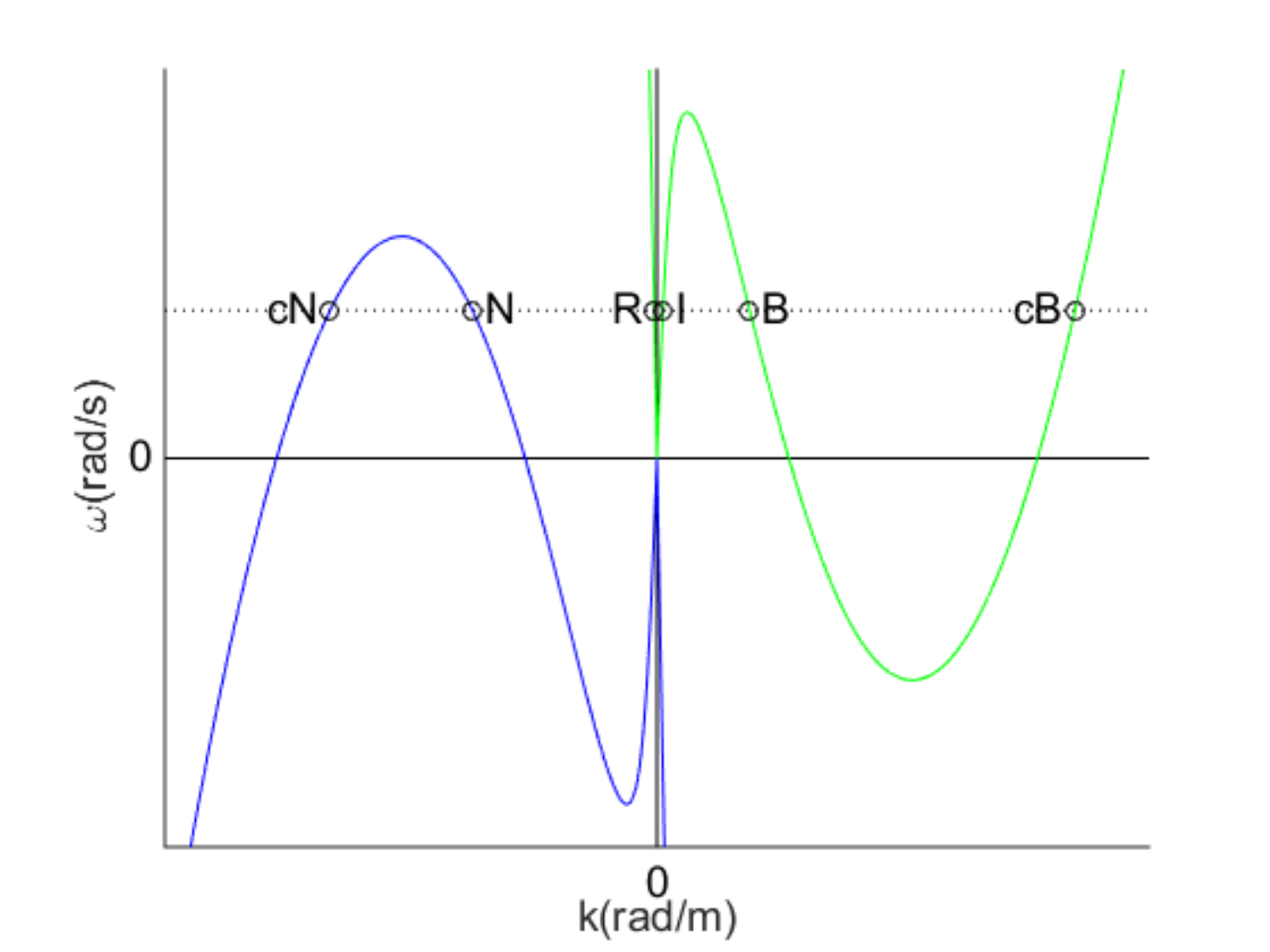}
\caption{Left: the horizons in the Fourier space (for the dispersion relation of capillary-gravity waves with $U<0$): white/black horizon (black diamond), blue/red horizon (white diamond), negative-white/negative-black horizon (gray diamond); the branches with a positive relative frequency are in green and the ones with a negative relative frequency are in blue. Right : the six solutions for a given frequency for a white hole configuration: incident wave (I), retrograde (R), blue-shifted (B), capillary positive (cB), negative (N), capillary negative (cN) modes.}
\label{horizon_solution}
\end{figure}

A dispersive horizon (namely a turning point) is such that the group velocity ($v_g=\frac{\partial \omega}{\partial k}$) is null for a certain angular frequency $\omega$: it corresponds to an extremum in the graphical representation of the dispersion relation (see the Figure \ref{horizon_solution} left). In presence of surface tension, both dispersive lengths mentioned above create either three or six horizons  depending on the configurations (half a bump/trough or a complete bump/trough). A non-dispersive white horizon of a white hole will degenerate into three dispersive horizons : the so-called white, blue and negative-white horizons \cite{NJP10,Como}. Conversely, a non-dispersive black horizon will also degenerate into three other dispersive horizons by time reversal symmetry: the so-called black, red and negative-black horizons. At a different position from the blocking points, we can have up to six solutions for the fixed frequency sent by, say, a wave-maker in a water channel (see the Figure \ref{horizon_solution} right).

\subsection{Critical periods : $T_c$ and $T_b$}

\begin{figure}[!htbp]
\includegraphics[scale=0.6]{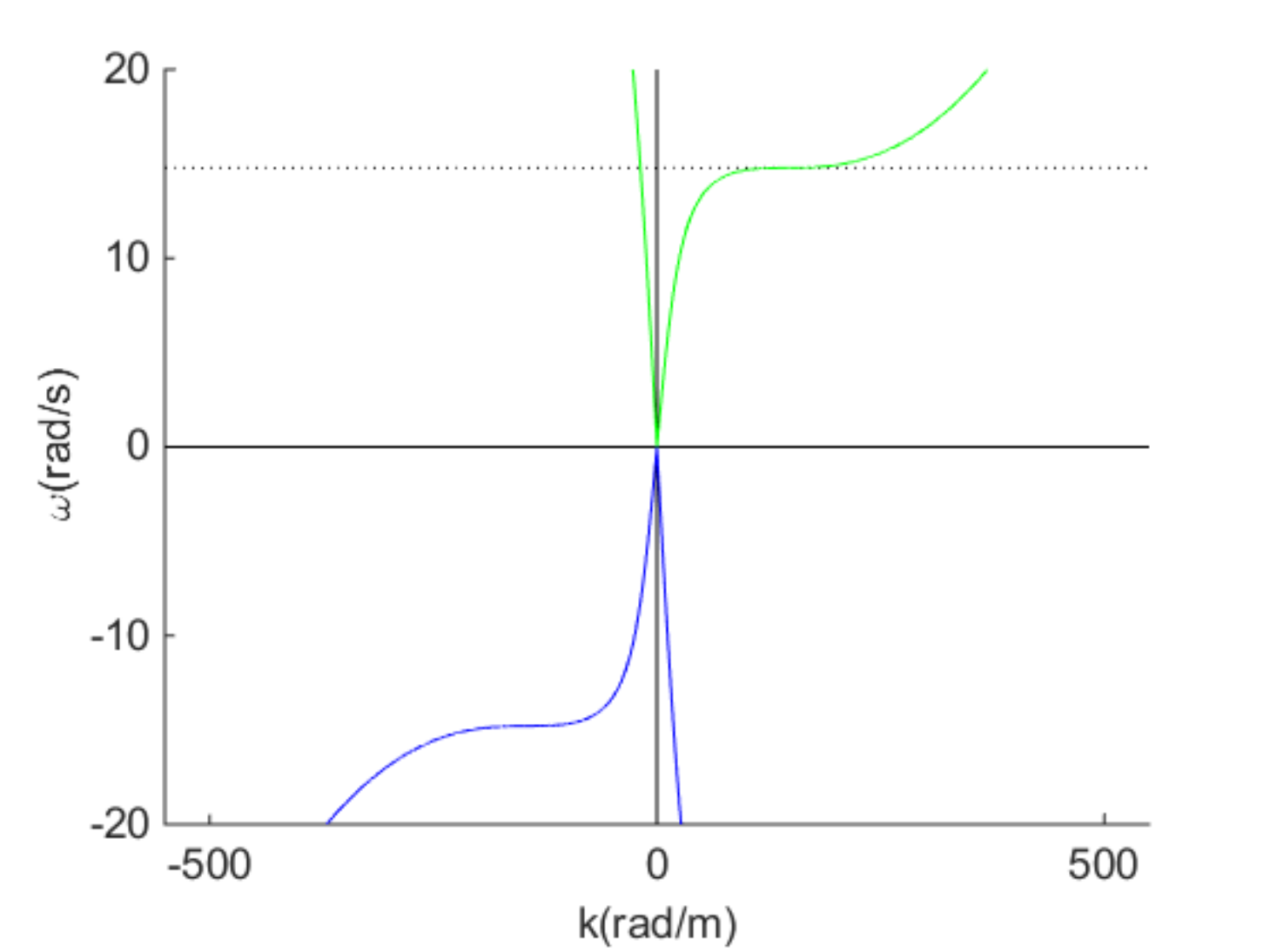}
\includegraphics[scale=0.6]{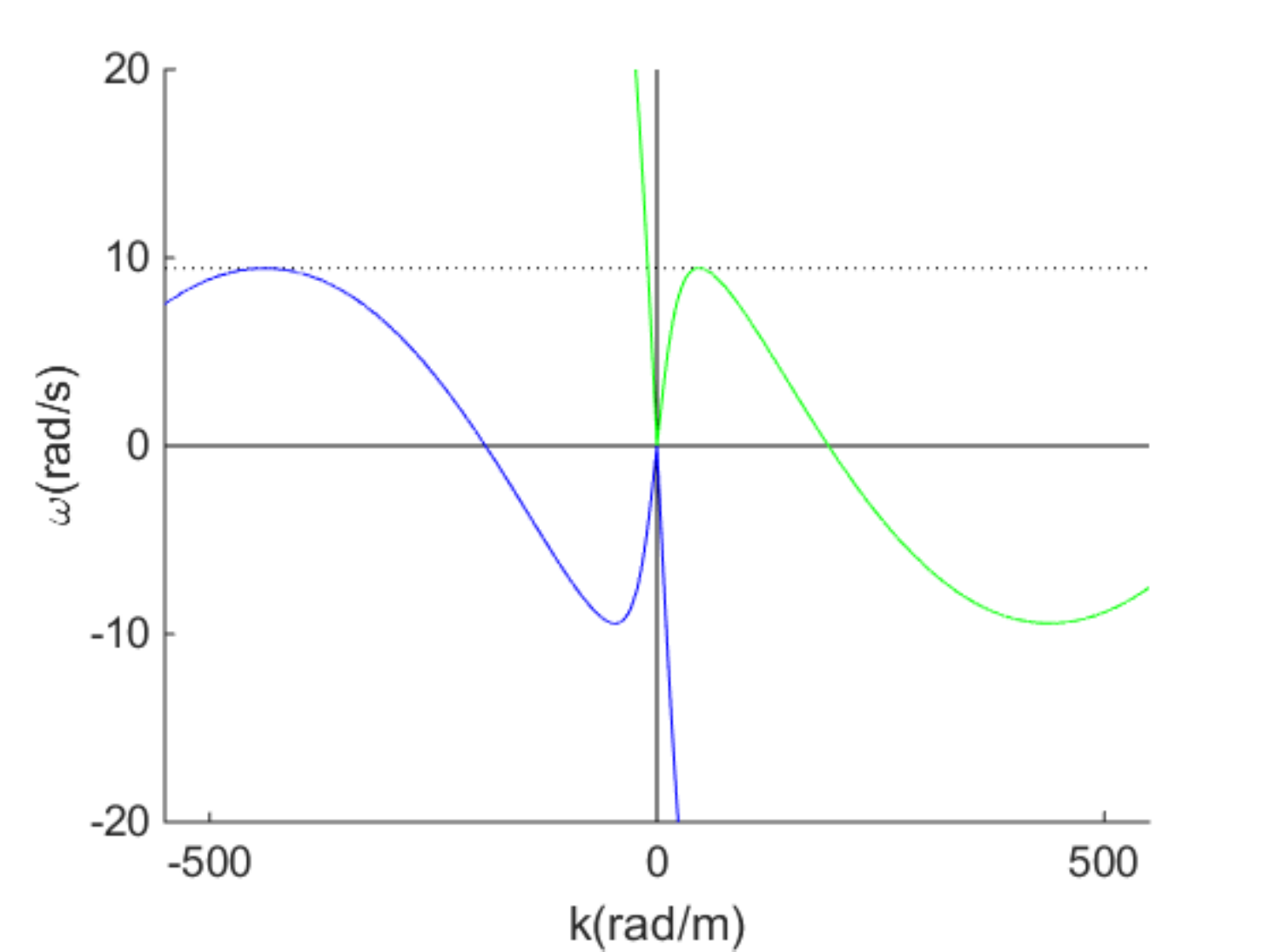}
\caption{Left: the dispersion relation with $U_c=-0.178m.s^{-1}$, $h=0.05m$ and $\omega_c=\frac{2\pi}{T_c}$ (black dot), $T_c=0.425s$. Right: the dispersion relation with $U_b=-0.255m.s^{-1}$, $h=0.05m$ and $\omega_c=\frac{2\pi}{T_c}$ (black dot), $T_c=0.647s$).}
\label{TcTb}
\end{figure}

The wave-packet ($k_0$ has a frequency $\omega_0$) is created in a constant velocity region with a subcritical velocity. Then, we follow its evolution in time. In \cite{NJP10, Como}, different regimes were distinguished (that depend on the period $T=\frac{2\pi}{\omega_0}$ of the incoming capillary-gravity wave-packet) separated by two peculiar periods $T_c$ and $T_b$.  When $T=T_c$ (see the Figure \ref{TcTb} left), both white and blue horizons merge and disappear for $T<T_c$. In that case, the incoming wave is so blue-shifted by the counter-current that it becomes a capillary wave which does not see anymore the white horizon: the so-called direct penetration scenario. This is similar to the behavior of phonons in a BEC with a Bogolyubov-type dispersion relation \cite{BLV, Como}. For $T=T_b$ (see the Figure \ref{TcTb} right), negative and white horizon are at the same position and for $T<T_b$, the conversion from an incident mode to a negative mode is impossible due to the absence of negative solution at the white horizon. In this work, we will not report simulations on this case ($T<T_c$ or $T<T_b$) which is already known in the literature but we will focus on the case where $T>T_b>T_c$ with the so-called double-bouncing scenario discussed later.

\section{Analogue dispersive bi-directional wormhole (direction : White$\rightarrow$Black)}

\begin{figure}[!htbp]
\includegraphics[scale=0.6]{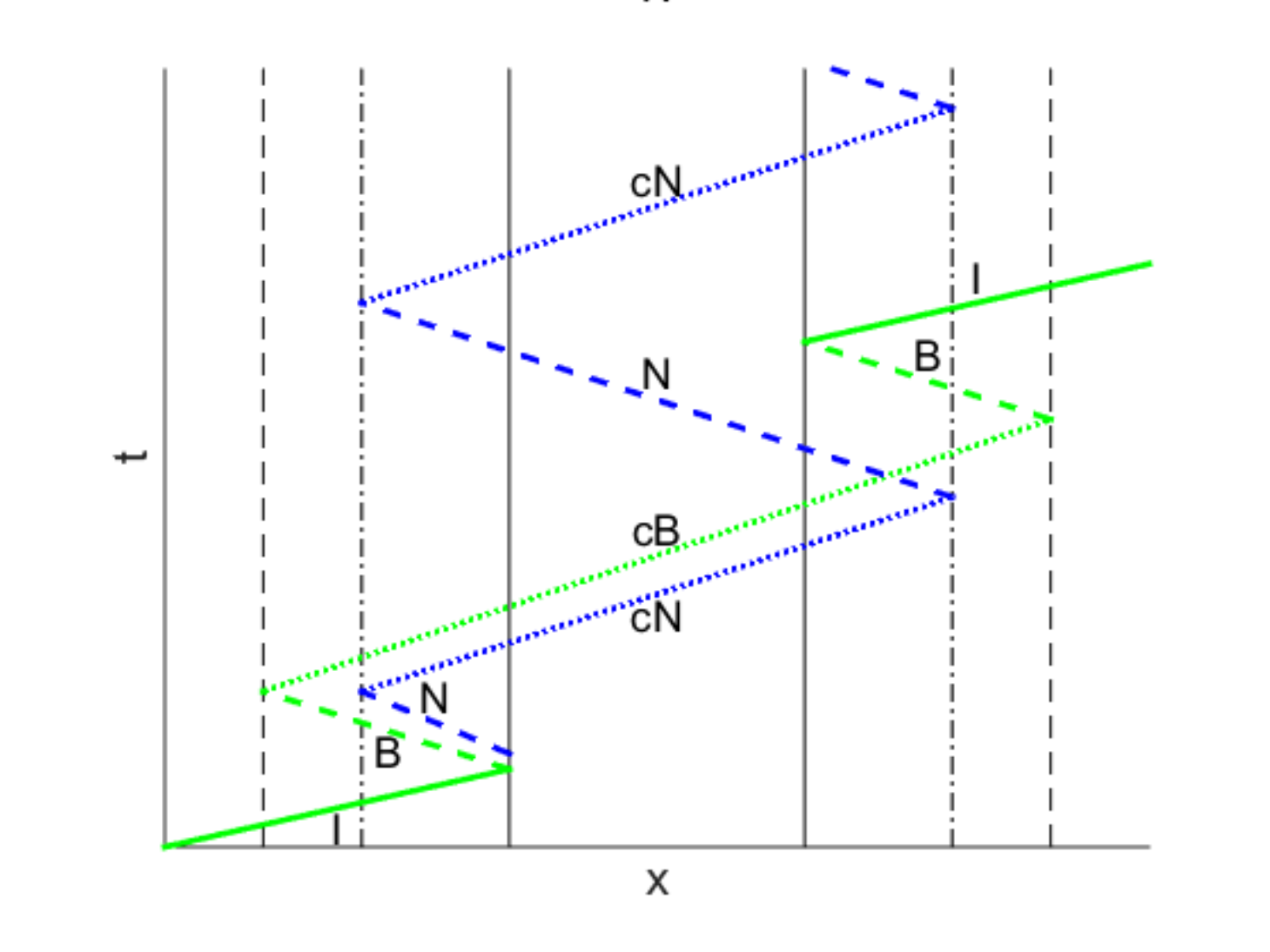}
\includegraphics[scale=0.6]{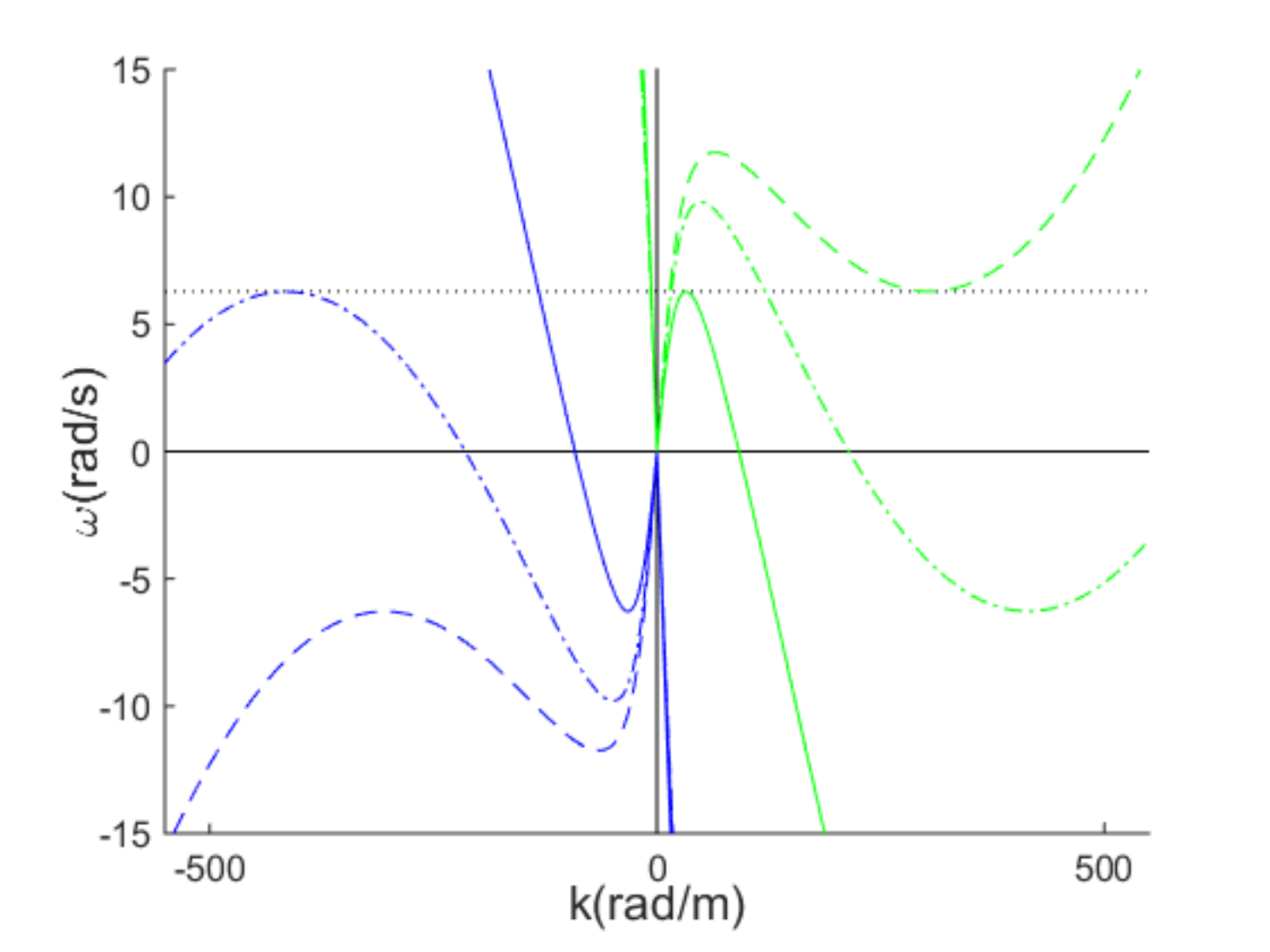}
\caption{Left: scheme of the main conversions in the travel - White$\rightarrow$Black -. The vertical lines represent the horizons positions, white(right)/black(left) horizons(continuous lines) where $U=-0.337m^2/s$, blue(right)/red(left)  horizons(dashed lines) where $U=-0.213m^2/s$, negative(right)/negative-black(left) horizons(dot-dashed lines) where $U=-0.248m^2/s$, $h=0.05m$, $T=1s$. Right: the corresponding dispersion relations at the different horizons for the corresponding velocities.}
\label{schemaWtoB}
\end{figure}

The wave packet $I$ is sent in a counter-current (see Figure \ref{velocity_field} right and \ref{schemaWtoB} left) from the left sub-critical region ($U=U_{min}$ and $x<x_1$). 

We recall that according to the theory and simulations described in Rousseaux et al. \cite{PRL09, NJP10, Como}, a long gravity wave propagating on a counter-current should be blocked at an analogue white hole horizon. An infinite blue-shifting is avoided since when the wavelength reaches the first dispersive scale (a combination of the angular frequency $\omega$ and the gravity field $g$, see \cite{PRL09} or the water depth in shallow water, see \cite{Como}), the incoming wave is either mode-converted into a so-called blue-shifting gravity wave whose wavelength matches the incoming wavelength at the blocking point in deep water or is blocked at turning point which depends on the incoming period, the water depth and the gravity field in shallow water. Unfortunately, the trans-Planckian problem is only displaced from the incoming to the blue-shifted gravity waves in deep water \cite{NJP10, Como}. The second dispersive length at a smaller scales (the capillary length), resolves the secondary trans-Planckian problem for the blue-shifted gravity waves by mode converting them towards smaller capillary waves at the "blue" horizon. Partners with negative relative  frequency in the current frame, necessary for Hawking radiation, are similarly mode-converted into negative capillary waves at the ``negative" horizon and cannot escape to infinity (similarly to the case of massive particles which may bounce back at a ``red" horizon \cite{PRD11}). To summarize, incoming gravity waves are mode-converted toward shorter capillary waves by bouncing twice successively at the white and blue horizons, and the shorter capillary waves enter into the normally forbidden white hole region, being ``superluminal" with respect to the ``super-critical" flow: the double-bouncing scenario \cite{Badulin, TM, NJP10, Como}. 

\begin{figure}[!htbp]
\includegraphics[scale=0.6]{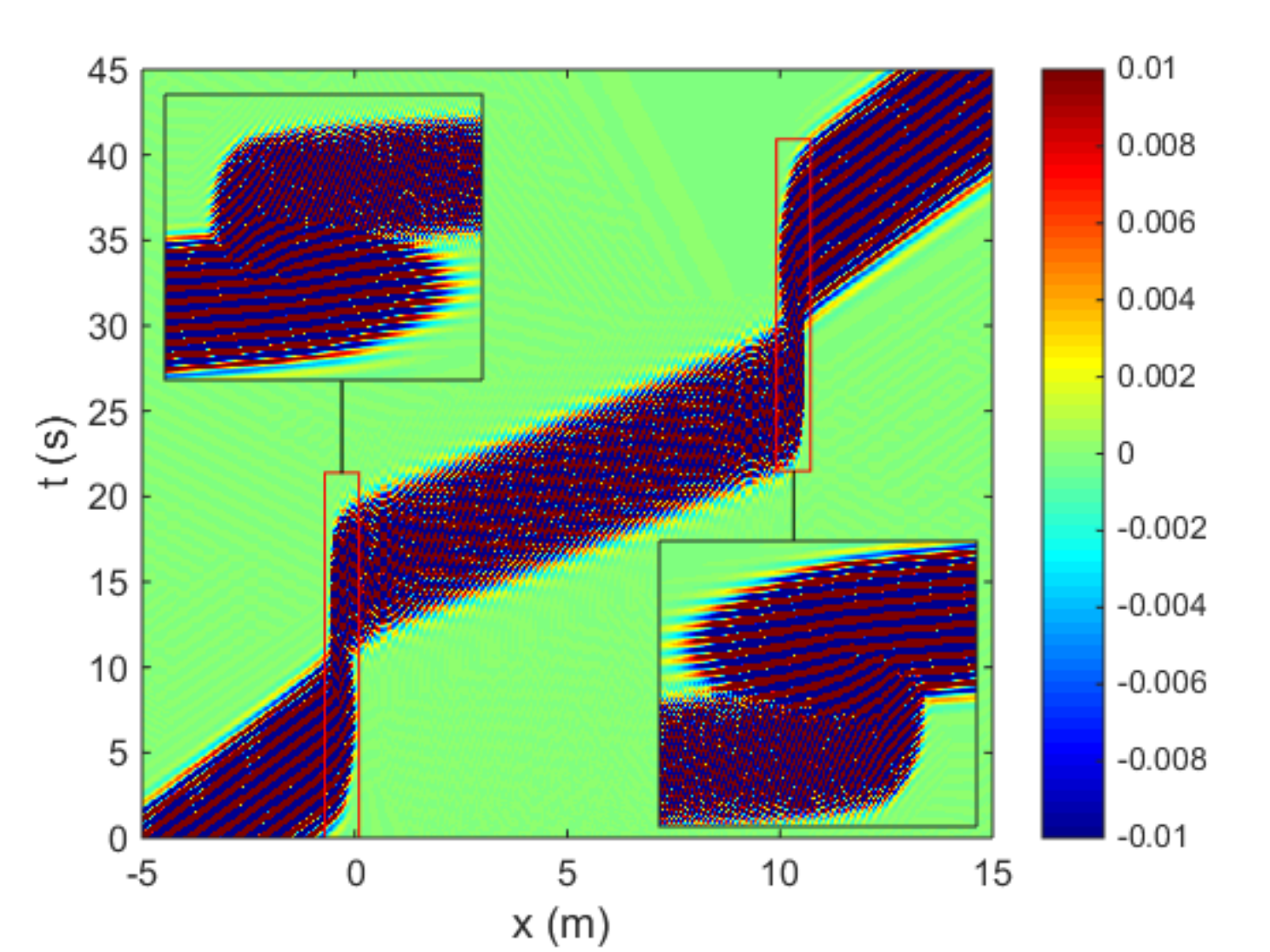}
\includegraphics[scale=0.6]{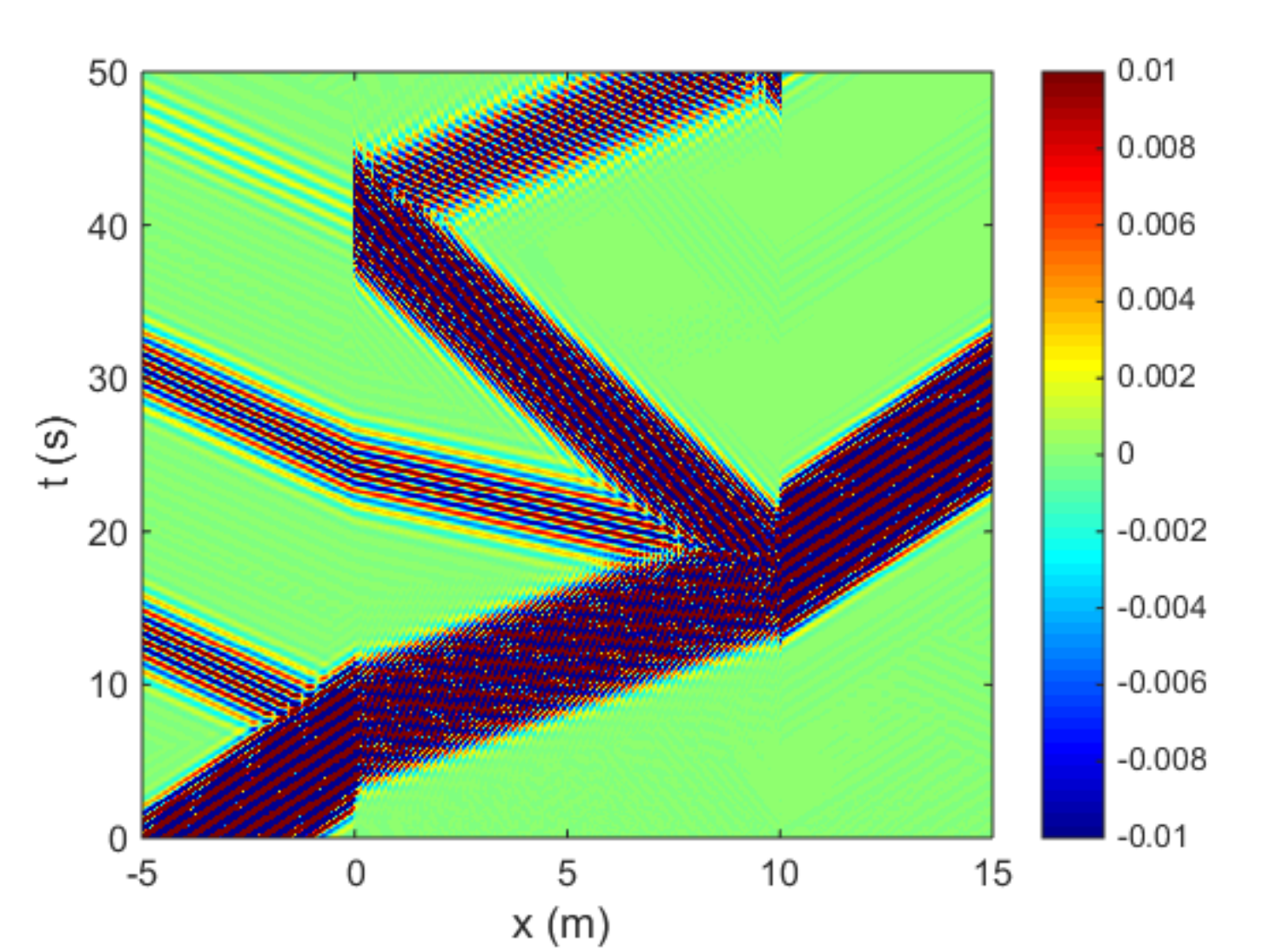}
\caption{Space-time diagram of the propagation of a Gaussian wave-packet in an analogue dispersive bi-directional wormhole (direction : White$\rightarrow$Black). The horizons are located roughly at $x_1=0m$ and $x_2=10m$. $h=0.05m$, $T=1s$, $U_{min}=-0.1m.s^{-1}$, $U_{max}=-0.9m.s^{-1}$. Left : $a_1=a_2=1.5m^{-1}$. Right : $a_1=a_2=20m^{-1}$}
\label{WtoB}
\end{figure}

\begin{figure}[!htbp]
\includegraphics[scale=0.82]{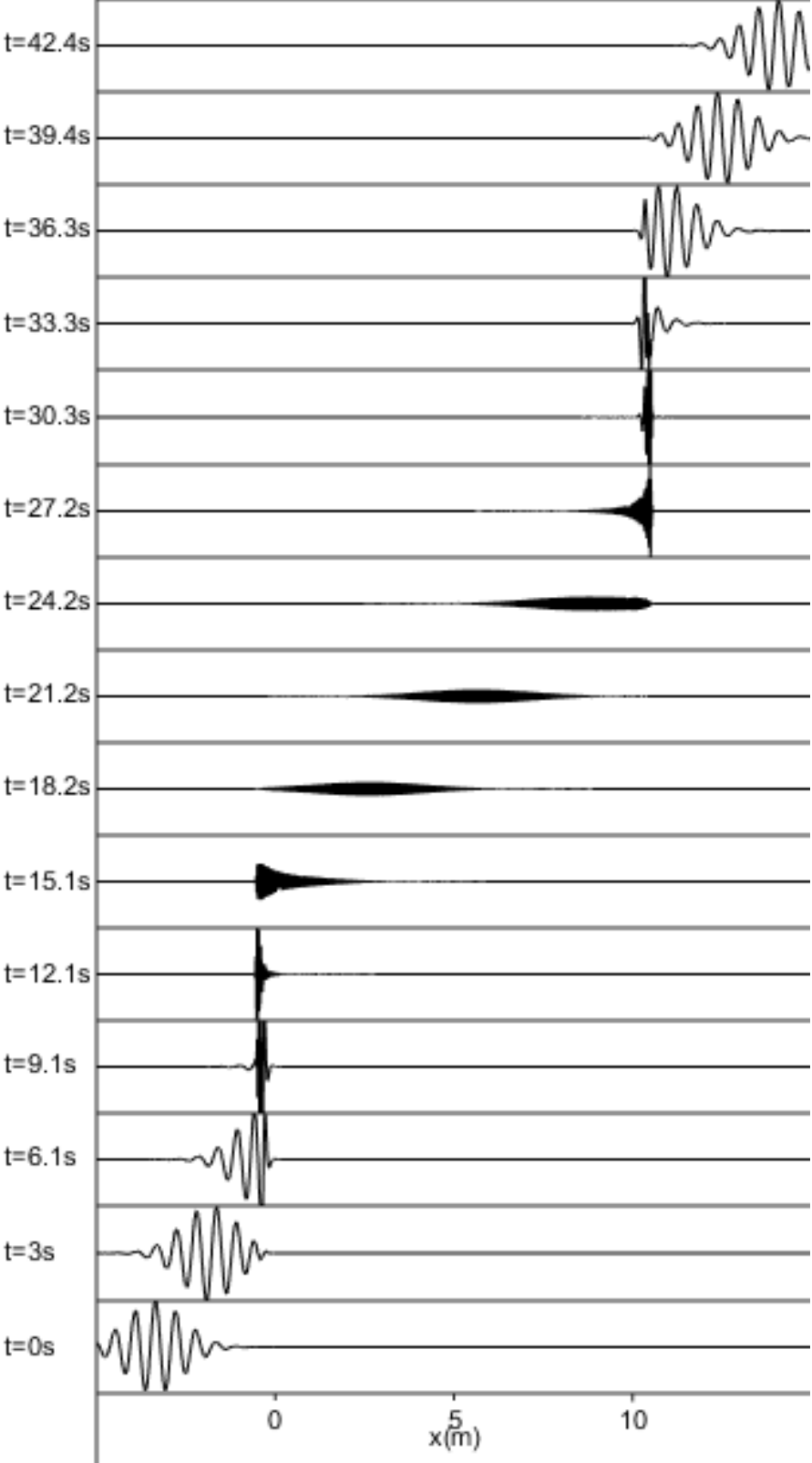} ~~~~~~~~~~~~
\includegraphics[scale=0.82]{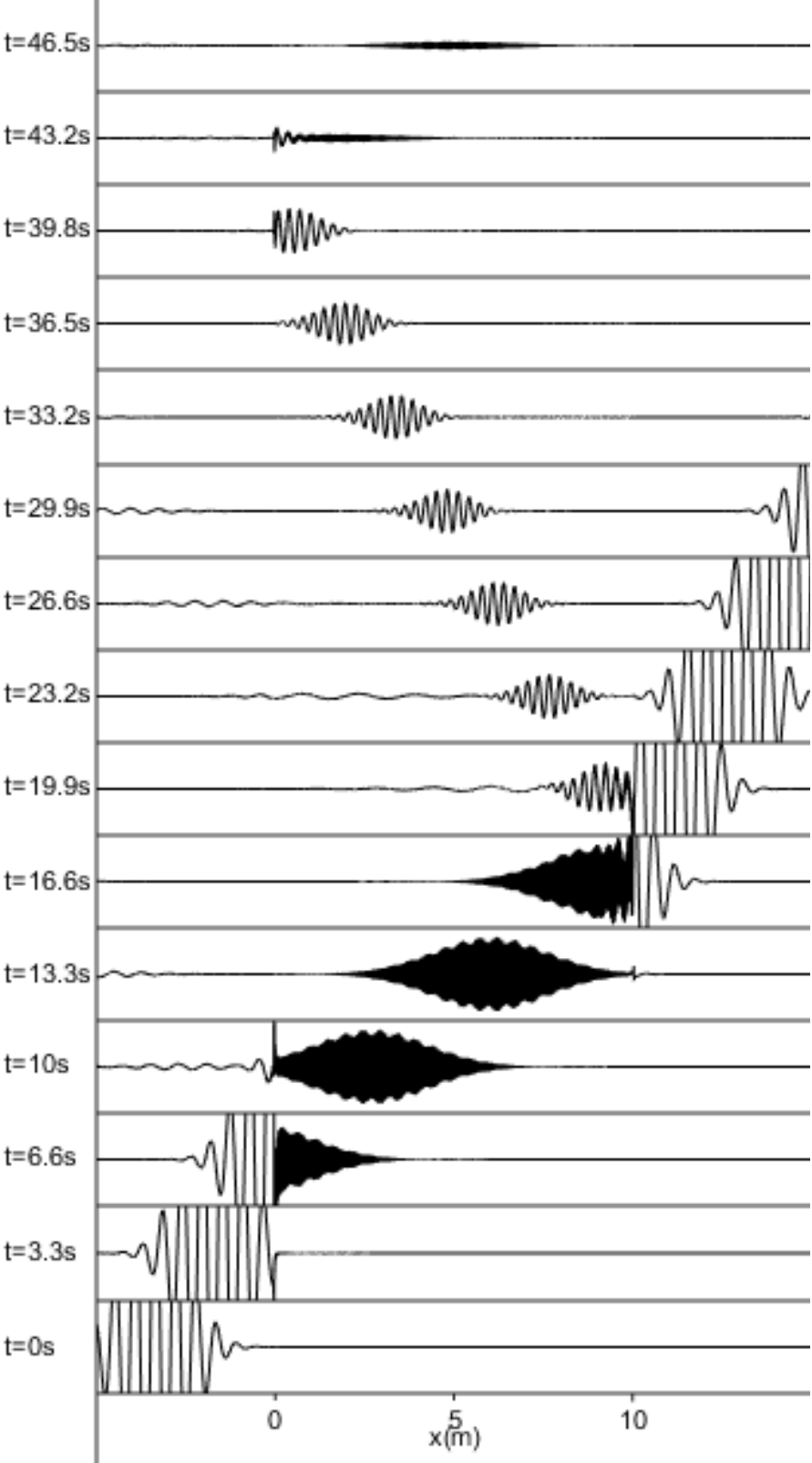} ~~~~~~~~~~~~
\caption{Left: successive snapshots of the propagation of a Gaussian wave-packet in an analogue dispersive bi-directional wormhole (direction: White$\rightarrow$Black) corresponding to the space-time diagram of the Figure \ref{WtoB} left (amplitude scale [-1:1]). Right: corresponding to the space-time diagram of the Figure \ref{WtoB} right (amplitude scale [-0.2:0.2]).}
\label{snapshotWtoB}
\end{figure}

Since the conversion to the different modes is controlled by the velocity gradient at the white horizon ($\frac{\partial U}{\partial x}|_{x^*}\sim a_1\cdot S$), we show two cases with different velocity gradients. Firstly, we use a small slope ($a_1=a_2=1.5m^{-1}$ see the Figures \ref{WtoB} left and \ref{snapshotWtoB} left), so that the conversion to the negative mode is negligible. We observe the path of the incident wave packet entering the white hole and leaving the analogue wormhole from the black hole side thanks to the double bouncing scenario at the four horizons (white-blue then red-black) created by the two dispersive lengths. Indeed, the capillary positive wave $cB$ generated by the double-bouncing in the white hole region follows the reverse path in the black hole region: it is "inversely" converted and red-shifted at the red horizon and re-converted at the black horizon as an incident wave.

Secondly, we focus on the conversion to the negative mode with a larger slope ($a_1=a_2=20m^{-1}$ see the Figures \ref{WtoB} Right and \ref{snapshotWtoB} right). The white, blue and negative horizons are very close spatially in this case. We observe the rebound of the couple negative $N$/negative capillary $cN$ wave between their respective horizons. At each rebound, parts of the wave packet energy are converted into incident and retrograde modes. The negative capillary wave $cN$ is blocked at the negative-black horizon and goes back as a negative wave $N$ to the negative-white horizon. This creates a recurrence $N-cN-N$ between this two negative horizons (see the Figures \ref{WtoB} right and \ref{snapshotWtoB} right): the black hole LASER effect. This phenomenon is observed here in numerical simulations for water waves with six horizons at play and it can be considered as a generalization of the previous proposals in Analogue Gravity featuring only two horizons.

\section{Analogue dispersive bi-directional wormhole (direction : Black$\rightarrow$White)}

\begin{figure}[!htbp]
\includegraphics[scale=0.6]{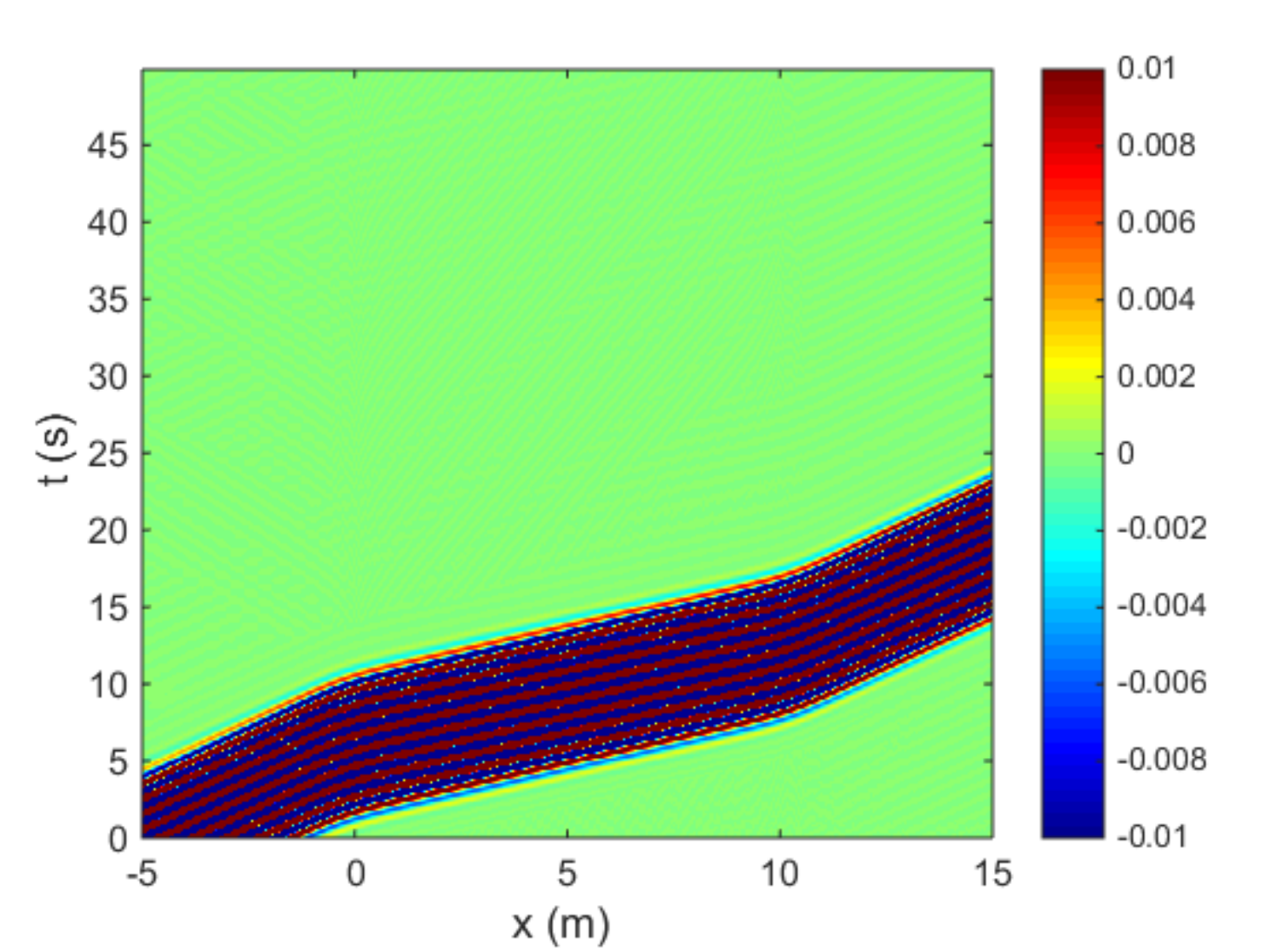}
\includegraphics[scale=0.6]{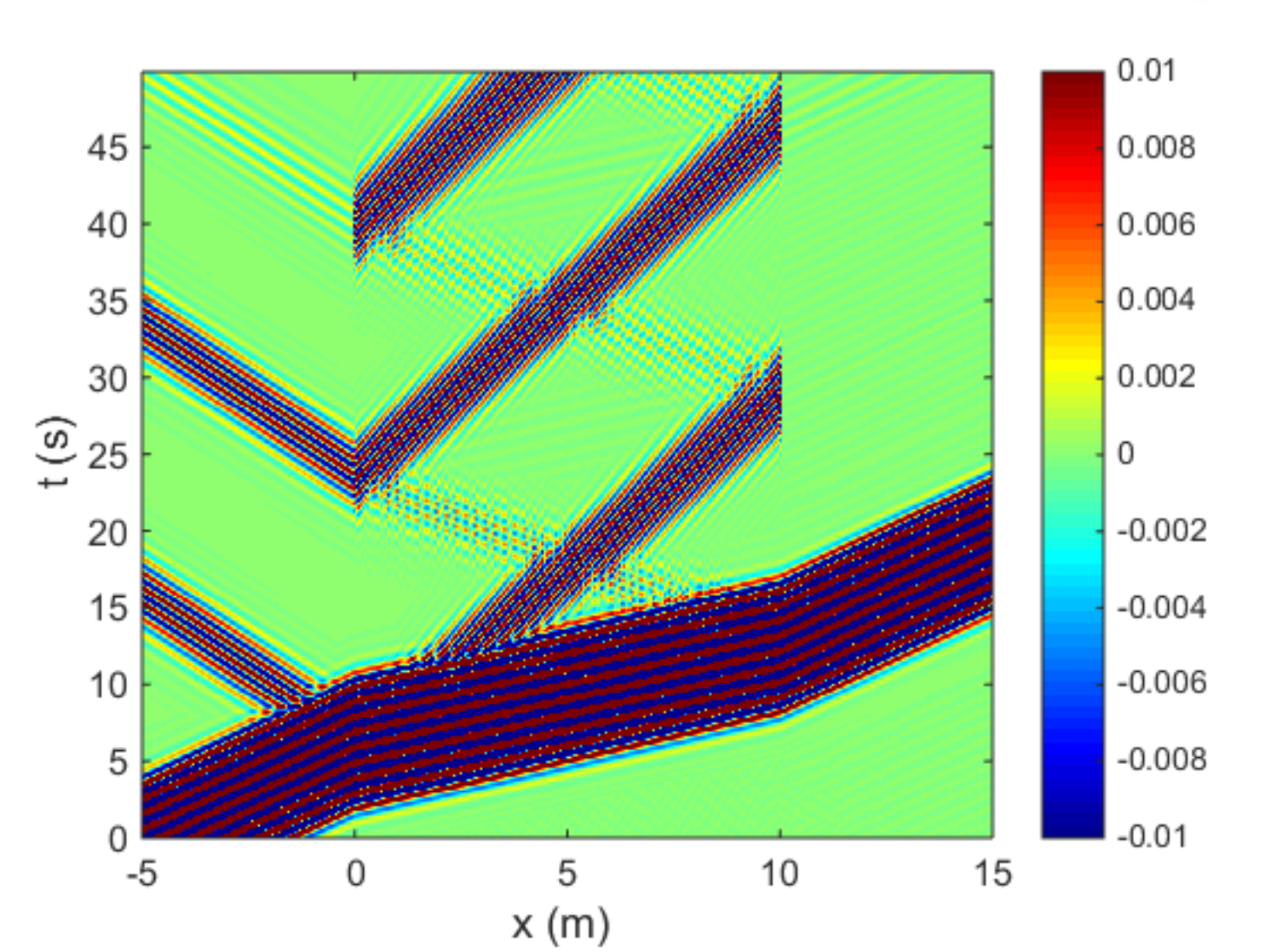}
\caption{Space-time diagram of the propagation of a Gaussian wave-packet in an analogue dispersive bi-directional wormhole (direction: Black$\rightarrow$White). The horizons are located roughly at $x_1=0m$ and $x_2=10m$. $h=0.05m$, $T=1s$, $U_{min}=0.1m.s^{-1}$, $U_{max}=0.9m.s^{-1}$. Left: $a_1=a_2=1.5m^{-1}$. Right: $a_1=a_2=20m^{-1}$}
\label{BtoW}
\end{figure}

\begin{figure}[!htbp]
\includegraphics[scale=0.82]{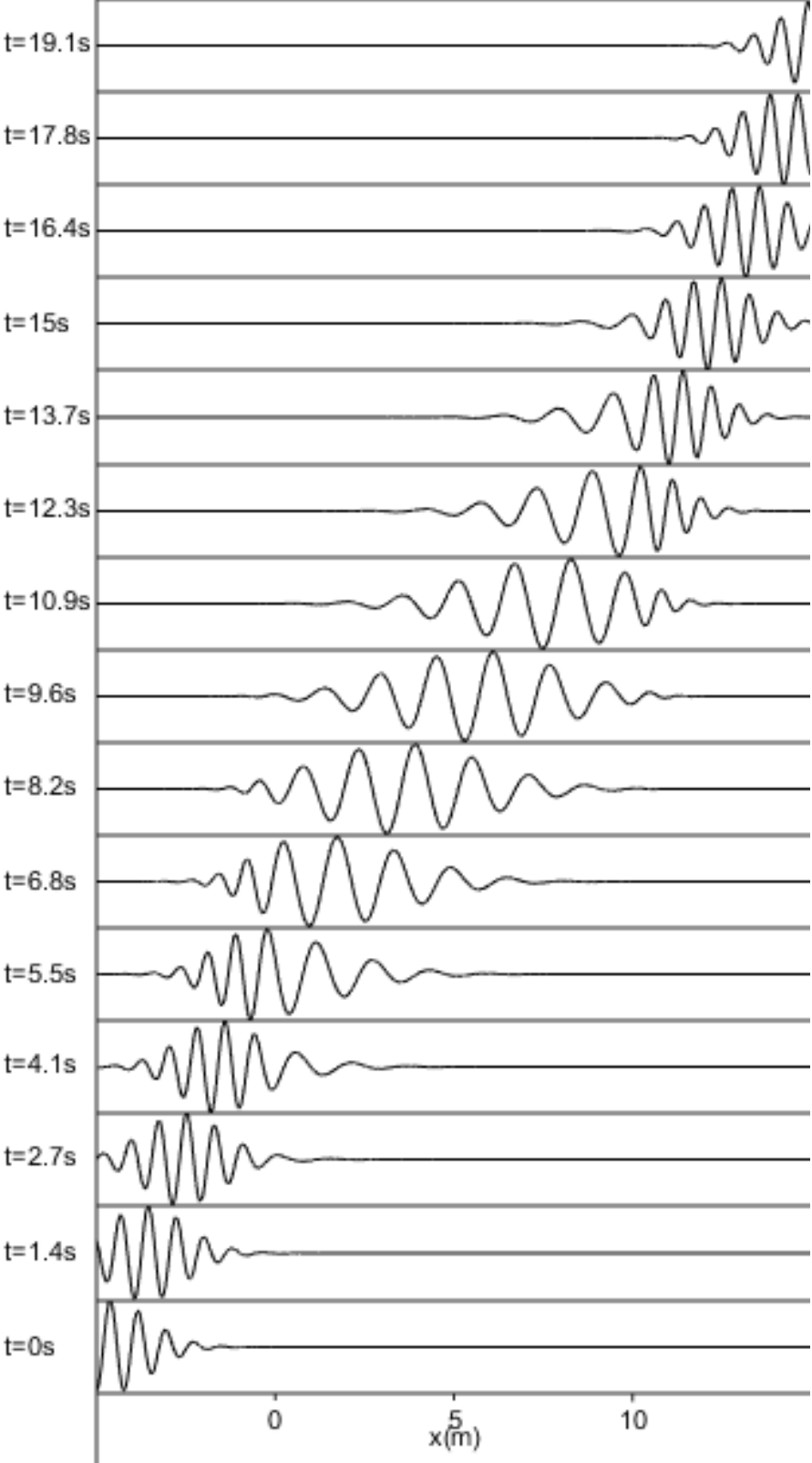} ~~~~~~~~~~~~
\includegraphics[scale=0.82]{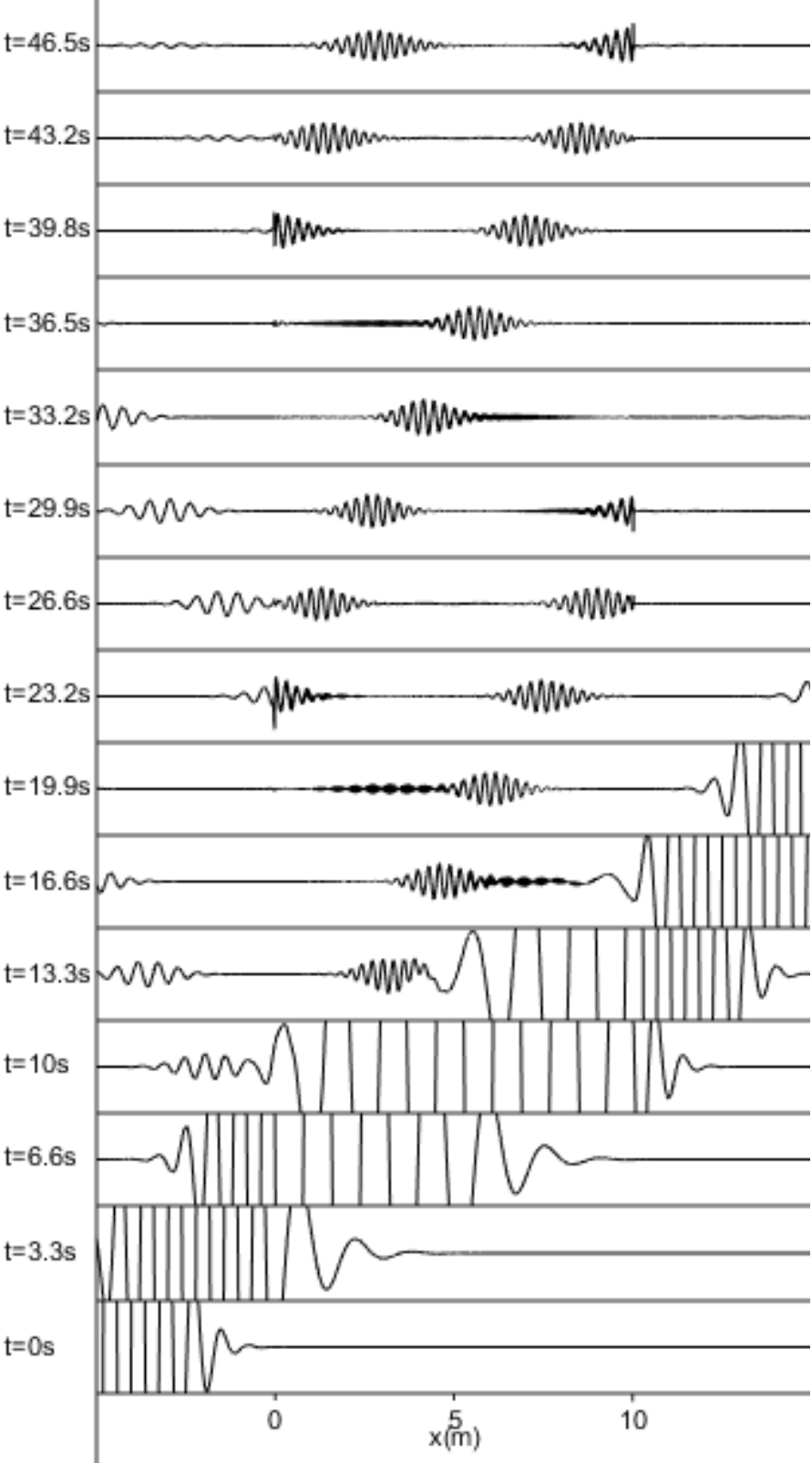} ~~~~~~~~~~~~
\caption{Left: successive snapshots of the propagation of a Gaussian wave-packet in an analogue dispersive bi-directional wormhole (direction: Black$\rightarrow$White) corresponding to the space-time diagram of the Figure \ref{BtoW} left (amplitude scale [-1:1]). Right: corresponding to the space-time diagram of the Figure \ref{BtoW} right (amplitude scale [-0.05:0.05]).}
\label{snapshotBtoW}
\end{figure}

In this case the wave packet is sent in the co-current direction - Black$\rightarrow$White - (see the Figure \ref{velocity_field} left) from the left sub-critical region ($U=U_{min}$ and $x<x_1$). If the slope is small ($a_1=a_2=1.5m^{-1}$ see the Figures \ref{BtoW} left and \ref{snapshotBtoW} left), the wave packet is just refracted by the velocity gradient but there is no conversion to other modes during the propagation. But if the slope is larger ($a_1=a_2=20m^{-1}$ see the Figures \ref{BtoW} right and \ref{snapshotBtoW} right), there is conversion without horizon at the point where the gradient is the stronger ($x_1$ and $x_2$) and we observe the same phenomenon as in the direction - White$\rightarrow$Black -. Here, the double bouncing scenario happens twice around each maximum of the velocity gradient.

\section{Conclusion}

We have shown numerical simulations of a LASER effect in the propagation of water waves in an analogue wormhole geometry (bump in the velocity profile). Six horizons are involved because of the existence of two dispersive scales contrary to the only scenario considered in Analogue gravity so far with two horizons and one dispersive scale. To test these predictions experimentally will be a challenge since we neglected two phenomena which will modify the current picture. First, also because of dispersion, a zero-frequency undulation (solutions of the dispersion relation $k\neq 0$ for $\omega=0$) will appear between both horizons as observed in the BEC case by Steinhauer \cite{Steinhauer}: this zero mode will be the background on which the converted waves will be superposed assuming that the effects stay linear. Finally, in the case of an analogue wormhole geometry, the small capillary waves will be, in practice, damped very rapidly as discussed in \cite{Mainardi} by the viscous dissipation on a distance shorter than the inter-horizon distance. A velocity profile where the inter-horizon distance is smaller than the dissipative length is required for the wormhole geometry and this implies additional theoretical investigations to assess the effect. For example, the complex frequency modes and correlation functions should be computed as was done in the BEC system \cite{Parentani}. Finally, the saturation of the LASER operation due to dynamical instabilities should also be investigated \cite{Florent}.\\

{\it Acknowledgements} : this research was supported by the University of Poitiers (ACI UP on Wave-Current Interactions 2013-2014), by the Interdisciplinary Mission of CNRS (PEPS PTI 2014 DEMRATNOS) and by the University of Tours in a joint grant with the University of Poitiers (ARC Poitiers-Tours 2014-2015). The french national research agency (ANR) funds the current work on the subject through the grant HARALAB (ANR-15-CE30-0017-04). We thank Scott Robertson for comments on a draft.

\end{document}